\documentclass[namedreferences,12pt]{mySolarPhysics}
\usepackage[optionalrh,linksfromyear,solaromanenum]{myspr-sola-addons} 
\usepackage{epsfig}            
\usepackage{graphicx}          
\usepackage{color}             
\usepackage[breaklinks]{hyperref}
\usepackage{breakurl}          
\ifx \doiurl    \undefined \def \doiurl#1{\href{http://dx.doi.org/#1}{\textsf{DOI}}}\fi
\ifx \adsurl    \undefined \def \adsurl#1{\href{http://adsabs.harvard.edu/abs/#1}{\textsf{ADS}}}\fi
\ifx \arxivurl  \undefined \def \arxivurl#1{\href{http://arxiv.org/abs/#1}{\textsf{arXiv}}}\fi

\newcommand{\aap}{{\it Astron. Astrophys.}}

\newcommand{\apj}{{\it Astrophys. J.}}

\newcommand{\apss}{{\it Astrophys. Spa. Sci.}}

\newcommand{\mnras}{{\it Mon. Not. Roy. Astron. Soc.}}

\newcommand{\pasj}{{\it Pub. Astron. Soc. Japan}}

\newcommand{\solphys}{{\it Solar Phys.}}

\newcommand{\ssr}{{\it Space Sci. Rev.}}

\def\degns{\ifmmode^\circ\else$^\circ$\fi}
\def\deg{\ifmmode^\circ\else$^\circ$\fi}

\def\gsim{\lower.4ex\hbox{$\;\buildrel >\over{\scriptstyle\sim}\;$}}
\def\lsim{\lower.4ex\hbox{$\;\buildrel <\over{\scriptstyle\sim}\;$}}

\begin{document}
\begin{article}
\begin{opening}
\title{Observations of  excitation and damping of  transversal  oscillation in coronal loops by AIA/SDO }
\author{  A.~\surname{Abedini}}
\runningauthor{A.~Abedini}
\runningtitle{Observations of  excitation and damping of  transversal  oscillation in coronal loops}
   \institute{Department of Physics, University of Qom,  Qom University Blvd Alghadir, P.O. Box  3716146611, Qom, I. R. Iran.
 \\
            email: \url{a.abedini@qom.ac.ir}
                   \\
            email: \url{a_abedini448@yahoo.com}
             }
\begin{abstract}
The excitation and damping of transversal coronal loop oscillations and
quantitative relation between  damping time, damping quality (damping time per  period), oscillation amplitude,  dissipation mechanism and the wake phenomena are investigated.
The observed  time series  data with the \textit{Atmospheric Imaging Assembly} (AIA)  telescope
on NASA's \textit{Solar Dynamics Observatory} (SDO) satellite   on  2015 March 2, consisting of  400 consecutive images
 with  12   seconds cadence in the 171 $ \rm{{\AA}}$ pass band is analyzed  for evidence of
  transversal  oscillations along the coronal loops by Lomb-Scargle periodgram.
In this analysis signatures of  transversal coronal loop oscillations that are damped rapidly were found
 with dominant oscillation periods in the range of $\rm{P=12.25-15.80}$ minutes.
 Also, damping times and damping qualities of transversal coronal loop oscillations
 at  dominant oscillation periods  are estimated in the range of $ \rm{\tau_d=11.76-21.46}$ minutes and $ \rm{\tau_d/P=0.86-1.49}$, respectively.
The observational results of this analysis show that damping qualities decrease slowly with increasing the amplitude of oscillation, but periods of oscillations  are not sensitive function of amplitude of oscillations.
The order of magnitude of the damping qualities  and damping times   are in good agreement with previous findings  and the theoretical prediction for damping of kink mode oscillations by dissipation mechanism.
Furthermore, oscillation of loop segments attenuate with time roughly as $t^{-\alpha}$ that
magnitude values of $\alpha$ for 30 different segments change from 0.51 to 0.75.
 \end{abstract}
\end{opening}
\section{Introduction}
Coronal seismology was first introduced  in the 20th century  by Uchida (\citeyear{Uchida1968},~\citeyear{Uchida1970}).
Then, it has been successfully used to probe of  physical parameters in the solar corona.
There are now many studies on   observations and  the theory of  solar  coronal seismology   (see, \textit{e.g.}
 $\rm{Erd\acute{e}lyi}$  \textit{et al.}~\citeyear{Erdelyi2003}; Wang ~\citeyear{Wang2004}; Aschwanden~\citeyear{Aschwanden2004},~\citeyear{Aschwanden2006}; Roberts~\citeyear{Roberts2004}; Nakariakov and Verwichte~\citeyear{Nakariakov2005}; Andries  \textit{et al.}~\citeyear{Andries2009}; Taroyan and $\rm{Erd\acute{e}lyi}$ ~\citeyear{Taroyan2009}; Ruderman and $\rm{Erd\acute{e}lyi}$ ~\citeyear{Ruderman2009}).
Observations  of solar corona with  space  telescopes such as  \textit{Hinode, Yohkoh, Transition Region And Coronal Explorer} (TRACE),  \textit{Solar Terrestrial Relations Observatory} (STEREO), \textit{Solar and Heliospheric Observatory} (SoHO) and  SDO  have helped  us to study in detail transverse oscillations  of coronal loops.
Today, there  are   many observations of  transverse motions of coronal loops.
For example, Nakariakov \textit{et al.}~(\citeyear{Nakariakov1999}),  Aschwanden \textit{et al.}~(\citeyear{Aschwanden1999})
and Verwichte \textit{et al.}~(\citeyear{Verwichte2004}) observed fast kink mode in coronal loops by TRACE.
Roberts \textit{et al.}~(\citeyear{Roberts1984}), Asai \textit{et al.}~(\citeyear{Asai2001}),
 Melnikov  \textit{et al.}~(\citeyear{Melnikov2005}) and Aschwanden \textit{et al.}~(\citeyear{Aschwanden2004}) observed fast sausage mode oscillations in radio wavelengths.
Morton \textit{et al.}~(\citeyear{Morton2011}) observed  symmetric sausage mode
in  a coronal loop  with \textit{Rapid Oscillations in the Solar Atmosphere} (ROSA) instrument.
Kohutova  and  Verwichte ~(\citeyear{Kohutova2016})  studied the kinematics and  oscillations   of the coronal rain
by, SOT/ \textit{Hinode,  Interface Region Imaging Spectrograph} (IRIS) and AIA/SDO.
They detected two distinct transverse oscillations regimes: large-scale oscillations
 and small-scale oscillations.
 The large-scale and  small-scale oscillations  were  exited  by a transient mechanism and  a continuous drive mechanism, respectively.
Goddard \textit{et al.}~(\citeyear{Goddard2016585})  estimated  the  physical parameters of  120 individual kink oscillations of coronal loops and  found that  period of oscillation scales with the coronal loop length, and the initial amplitude of oscillation is  determined by  the initial  displacement of coronal loop.
Also,  they found that the magnitude  of  kink speed of coronal loops are  in the range $\rm{C_k = (800-3300)~km s^{-1}}$.
Nakariakov \textit{et al.}~(\citeyear{Nakariakov1999}), Aschwanden and  Schrijver ~\citeyear{Aschwanden2002},
 and Aschwanden \textit{et al.}~(\citeyear{Aschwanden2002})
observed an important characteristic of  transversal coronal loop oscillations that the amplitude
of oscillations were damped strongly.
A large number of theoretical studies investigating
 the damping time  of the  transverse oscillations in a coronal loop  have  focused on the effects of phase mixing, resonant absorption, gravitational stratification, magnetic
field divergence (see, \textit{e.g.} Aschwanden \textit{et al.}~\citeyear{Aschwanden2003}; Safari \textit{et al.}~\citeyear{Safari2007}; Ebrahimi and Karami ~\citeyear{Ebrahimi2016}). In general, phase mixing and resonant absorption
are shown to be the main physical   mechanisms for damping  of
 the standing  transverse  oscillations of coronal loops.
The majority of previous studies on transversal coronal loop oscillations assumed that coronal loop is in a  stationary state.
In addition, dynamics of  a coronal loop  mainly  were determined  by the normal modes (see,\textit{e.g.}~ Edwin and Roberts~\citeyear{Edwin1982};~\citeyear{Edwin1983}; Spruit~\citeyear{Spruit1982}; Cally~\citeyear{Cally1986}, ~\citeyear{Cally2003}).
However,  the normal modes of  a coronal loop are not the full picture of the loop dynamics.
Only a few limited studies have been done about the time-dependent problem
of the excitation and damping of coronal loop oscillations.
For example, Murawski and Roberts~(\citeyear{ Murawski1993a}) studied the
time evolution of  driven waves in coronal loops which were  approximated by smoothed slabs, and  they  investigated some  properties  of sausage and kink  modes of these loops. Also, they studied the temporal signatures of impulsively generated fast waves in the solar coronal loops by means of numerical simulations (\citeyear{Murawski1993b},~\citeyear{Murawski1993c}).
Rae and Roberts~(\citeyear{Rae1982}), Uralov ~(\citeyear{Uralov2003}) and Terradas \textit{et al.}~(\citeyear{Terradas2005}) investigated the role of a traveling pulse generated by a localized event on coronal loop oscillations.
These authors  claimed that the wake of the propagating pulse  could be responsible for decay of the observed oscillations.
Many continuous observations of coronal loops with  satellite telescopes reveal that the transversal coronal loop oscillations are accompanied by a nearby flare  in an active region
(see, \textit{e.g.} Schrijver and Brown ~\citeyear{Schrijver2002}; Hudson and Warmuth~\citeyear{Hudson2004}; Aschwanden \textit{et al.} ~\citeyear{Aschwanden1999},~\citeyear{Aschwanden2002}; Aschwanden and Schrijver ~\citeyear{Aschwanden2011};  Nistic\`{o} \textit{et al.}~\citeyear{Nistico2013}; Verwichte \textit{et al.} ~\citeyear{Verwichte2013}; Anfinogentov \textit{et al.} ~\citeyear{Anfinogentov2015};  Hindman and Jain~\citeyear{Hindman2014}; Pascoe \textit{et al.} ~\citeyear{Pascoe2016}).
As a result, some authors claimed that excitation and damping of transversal
oscillation of coronal loops may  be caused  by the wake of a traveling disturbance from nearby flares (see, \textit{e.g.} Terradas \textit{et al.}~\citeyear{Terradas2005}).
Recently, Zimovets and  Nakariakov (\citeyear{Zimovets2015}) studied the excitation  of kink oscillations of coronal loops  by  flares and lower coronal eruptions/ejections (LCEs) of some unstable configuration of coronal magnetised plasma, such as  a  filament, a system of coronal loops, a magnetic flux rope and a  coronal mass ejection from a nearby flare site by analyzing  169 kink oscillation loops in  58 events.
In this analysis,  it was established that  nearby lower coronal eruptions is the most probable mechanism
 for exciting  decaying kink oscillations of coronal  loops compared to a blast shock wave  from  a flare.
In this paper,  I attempted to provide more insight into
the  excitation,  damping  and temporal evolution of   transversal  oscillations in solar coronal loops, and also  made an attempt  to deduce a quantitative relation between  damping time, damping quality, oscillation amplitude,  dissipation mechanism and the wake phenomenon that is produced by a traveling disturbance in solar coronal loops.
For this purpose, the nature of  transversal coronal loop oscillations (as observed with AIA/SDO) is analyzed by histogram equalization  technique and Lomb-Scargle periodgram algorithm that  shows a clearly better detection accuracy and efficiency   for analysis of noisy time series  data.
The magnitude values of physical parameters of transversal coronal loop oscillations such as
period of oscillation, decay time and damping quality are extracted
in 171 $\rm {\AA}$ passband. Moreover, both  observationally and theoretically
 the quantitative dependence of damping of  transversal coronal loop oscillation on dissipation
  mechanisms and the wake phenomenon  are studied.
The rest of this paper is organized as follows.
In Section 2  the observations are presented, in  Section 3
 the method of data analysis are described, in  Section 4   the physical parameters of  transversal coronal loop oscillation  are estimated, and   the theoretical considerations are given in Section 5.
Finally, discussion and conclusions are presented in  Section 6.
\section{Observations}
The data used for  this study are  EUV  images of the solar coronal loops, taken by AIA/SDO 171$\rm {\AA}$.
The initial raw images had to be   cleaned and calibrated  before using.
The images  used here are at level 1.5. For  images  with level 1.5,  co-alignment, flat-fielding,vignetting, filter  and bad-pixel/cosmic-ray corrections have already been  applied. Also, these images have been centered, rotated so that Sun's North Pole is  up in the image  and  rescaled down to  a plate scale of  0.6 arcsec/pixel. Additionally, the differential rotation effect is also corrected.
 \begin{figure}    
        \centerline{\includegraphics[width=.9\textwidth,clip=]{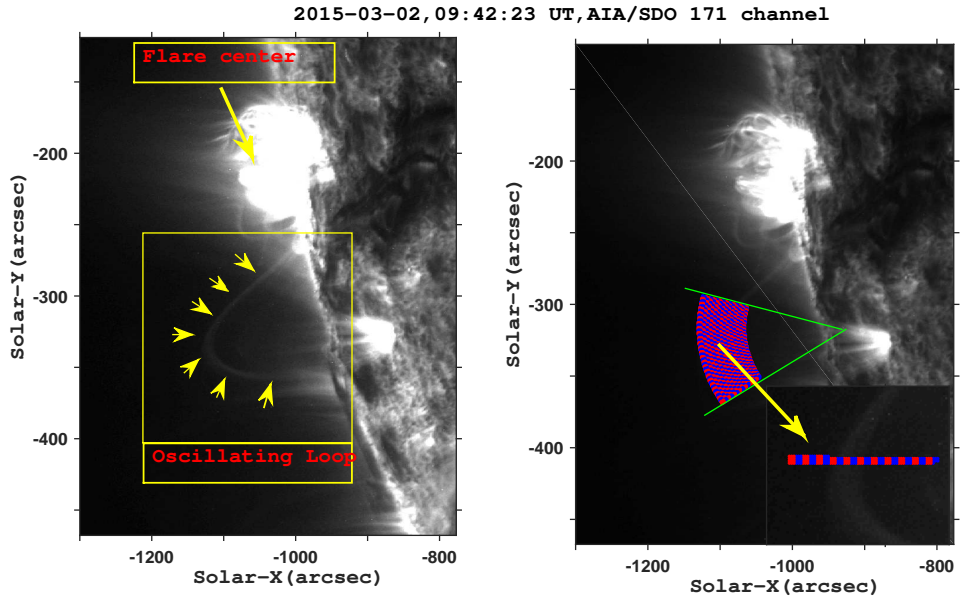}
              } \caption{A snapshot of the flare and near-by loop, like a semi circle, observed by AIA/SDO in 171$\rm {\AA}$ on 2015 March 2, 09:42:23 UT. The locations of the flare and loop are shown with yellow arrows in the left panel. To obtain the  transversal coronal loop oscillations, the part of area between two circular arcs containing the loop is subdivided into 30 sectorials that  are perpendicular to the visible loop strands. These sectorial areas are then joined by successive cross sections, three pixels wide (see one of these sectorials (zoomed) in the bottom of the right panel).}
                              \vspace{+0.01\textwidth}
   \label{fig1}
   \end{figure}
The data under analysis here is taken on 2015 March 2, from 09:00:00 UT until 10:30:00 UT
and it consists of a series of $1100\times600$ pixels, sun-centered, subfield images on the 171$\rm {\AA}$ (Fe IX) passband with a time cadence of 12 seconds.
The continuous  observations  of the selected  active region loop  and its environment monitored with  Helioviewer, JHelioviewer, \textit{Large Angle and Spectrometric Coronagraph} (LASCO) onboard the SoHO,  CME catalogue
and  CACTus CME list show that  there is a  flare and a LCE  in the same parental active region.
The solar flare that has been  located at position $\rm{ (x_{flare}, y_{flare})=(-960'',-192'')}$  near the loop   is started around 2015 March 2, 09:09:00 UT and  ended around 2015 March 2, 09:31:12 UT with
a peak flux  of 1182.3  $\rm{erg/(cm^{2}s)}$ that is occurred at 09:17:12 UT.
Also, the  LCE   that has been  located at position $\rm{ (x_{LCE}, y_{LCE})=(-1006.8'',-195.6'')}$  near the loop
 started around 2015 March 2, 08:21:07 UT and  ended around 2015 March 2, 09:01:07 UT.
The left panel of Figure \ref{fig1} shows a snapshot of the flare and a nearby loop in 171$\rm {\AA}$ on 2015 March 2, 09:24:23 UT that we are analyzing in this paper.
The locations of the flare and loop are shown with yellow arrows.
To obtain the  transversal coronal loop oscillations, the part of area between two circular arcs containing the semi circle-like loop is subdivided many distinct  sectorials that are perpendicular to the visible loop strands (Figure \ref{fig1},  right panel). These sectorial areas are then joined by successive cross sections, three pixels wide. Also,
one of the sectorial segments is zoomed in the lower part of the right panel.
Figure \ref{fig1} shows thirty distinct sectorial areas that are  analyzed
in detail   in the next sections.
\section{Data Analysis}\label{Dataa}
In order to extract  the oscillatory nature of transversal coronal loop oscillation,
the histogram of images is equalized into histogram of first image
by Image Processing Toolbox of Matlab. Histogram equalization is a technique for adjusting image intensities to enhance contrast.
\begin{figure}
   \centerline{\includegraphics[width=.9\textwidth,clip=]{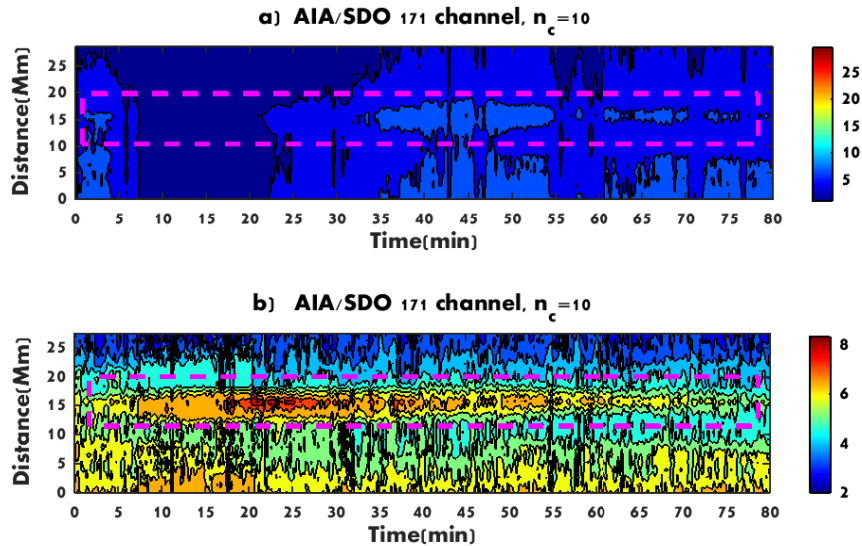}
              }\caption{Typical examples of time$-$distance  plot of original intensity (a)
           for sectorial 10 are shown in the \textit{top panel} before histogram equalization. Enhanced time$-$distance plot of intensity (b) for sectorial 10 are shown in the  \textit{bottom panel} after histogram equalization. Location of loop enclosed by a \textit{red rectangle} in the panels.}
           \vspace{-0.01\textwidth}
  \label{Fig2}
\end{figure}
Typical examples of time$-$distance plots of original
 (a) and histogram equalization  (b) intensities time  series  of images  for sectorial 10
are shown in the top and bottom panels of Figure \ref{Fig2}, respectively.
 Location of the loop enclosed by
a red rectangle in the panels. The part of area between two circular arcs containing the semi circle-like loop is subdivided into 30 sectorial areas that are  perpendicular to the visible loop strands. Moreover, each of these sectorial areas are joined by successive cross sections, about three pixels wide.
 The mean intensity of cross sections as  function of time is calculated by  dividing the total intensity of cross sections into  the number of pixels at each cross sections.
 A background intensity must be subtracted  from the original intensity of cross sections
   to enhance the contrast of  time$-$distance  plot and loop axis displacement.
 Different enhancement algorithms are used to remove the background for an improved visualization of
  transversal displacements of the loop axis  in a time$-$distance plot
 (see, \textit{e.g.} Aschwanden and Schrijver~\citeyear{Aschwanden2011}; Yuan and Nakariakov ~\citeyear{Yuan2012}; Threlfall \textit{et al.} ~\citeyear{Threlfall2013}; Abedini ~\citeyear{Abedini2016}).
Here, following Aschwanden and Schrijver~(\citeyear{Aschwanden2011}), a running-minimum difference in the form
\begin{eqnarray}
&&i(n_c,s_k,t_i)=I(n_c,s_k,t_i)-I_b,\\ \nonumber
&&I_b={\rm{min}}[I(n_c,s_{k},t_{i-j}),...,I(n_c,s_{k},t_{i+j})],
\label{iminrun}
  \end{eqnarray}
is  used to remove the background intensity of the time$-$distance plots.
This running-minimum  algorithm subtracts a running-minimum obtained within a time
interval with a length of $2j$ from  intensity time  series of macropixels that are located in the time$-$distance plots of  the sectorial areas.
 Where $n_c$ is the number of sectorial areas,  $s_k$ is  the location of the $k$th  macropixel at a specific sectorial area,
 $t_i$ is time of ith frame and $t_{i-j},...,t_{i+j} $ represent a time interval of $2j$ frames that  symmetrically are placed around $i$th frame in every time slice  ($n_c=1,k=7,i=2$ corresponds to the seventh macropixel along the sectorial 1 at $t=12s$). The width of selected arcs on the loop are varied between three to six macropixels.
An appropriate background  is subtracted from the original intensities. Sufficiently enhanced time$-$distance plots of sectorial areas are found by setting the $j=9,...,12$.
For example, time$-$distance plots of normalized background-subtracted intensities after histogram equalization ($i_n(s_k, t_i)=i(s_k, t_i)/max[i(s_{1},t_i),...,i(s_{end},t_i)]$) for thirty  sectorial areas by running-minimum  algorithm  are shown in Figures \ref{fig3} and \ref{fig4}.
 In order to improve visualization of  transversal  loop oscillation,  low oscillation periods ($\rm{P}\leq 36 s$) are filtered from intensities  after  running-minimum subtraction in the thirty time$-$distance plots.
The time$-$distance plots of intensities along the sectorial areas  show that  transversal coronal loop oscillations are occurred in the range of 13 to 43 minutes.
Clearly, the enhanced time$-$distance plots of sectorial areas are found by setting the $j=9,...,12$.
From Figures \ref{fig3} and \ref{fig4} it can be seen that the  histogram equalization technique has  enhanced
the contrast of transverse displacements of loop compared
to the previous  studies (see, \textit{e.g.} Aschwanden and Schrijver~\citeyear{Aschwanden2011}; Goddard \textit{et al.}~\citeyear{Goddard2016585}). Moreover, in these figures damping  and the start times  of transverse displacements  are clearly visible.
\begin{figure}
   \centerline{\includegraphics[width=.9\textwidth,clip=]{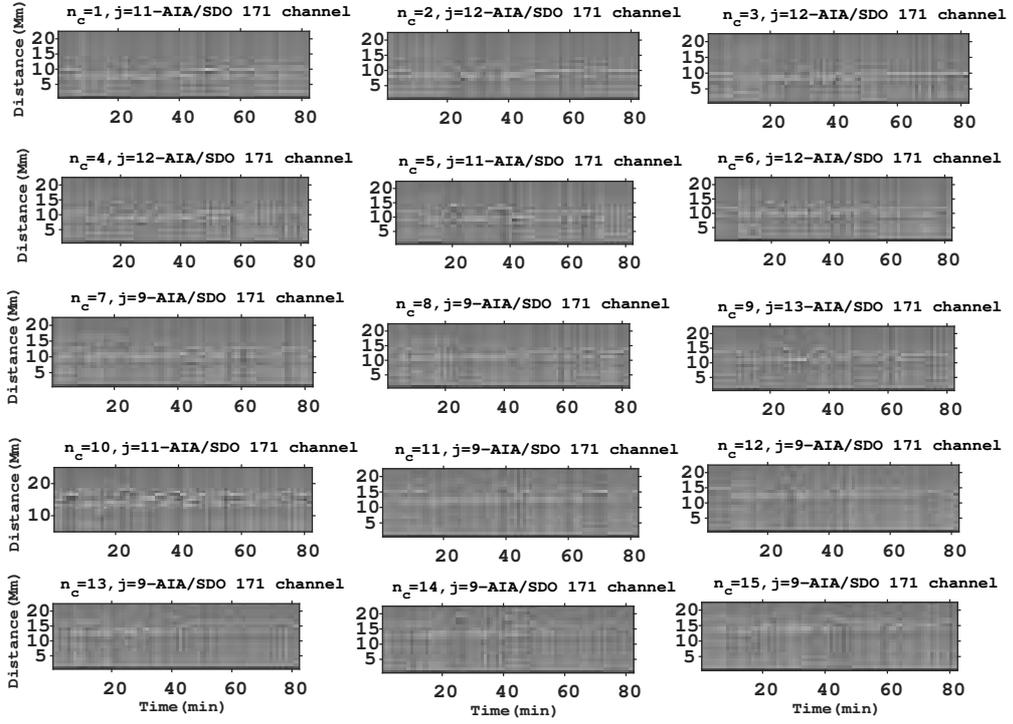}
              }\caption{ Running-minimum algorithm to visualize the  transversal coronal loop oscillations in  time$-$distance plots of  intensities ($i_n(s_k, t_i)=i(s_k, t_i)/max[i(s_{1},t_i),...,i(s_{end},t_i)]$)   in 171$\AA$ channel of AIA for the sectorial areas 1-15 (time range is 2015 March 2, 9:00:00-10:30:00 UT).
              In order to improve visualization of transverse displacements of loop   low oscillation periods ($\rm{P}\leq 36~s$) are filtered  from intensities  after  running-minimum subtraction in  the time$-$distance plots.
           Clearly, transversal coronal loop oscillations is occurred in the panels in the range of 15 to 45 minutes.
         The number of sectorial area ($n_c$) and magnitude value of $j$ are
         shown on  top of each panel.}
           \vspace{+0.01\textwidth}
     \label{fig3}
   \end{figure}
\begin{figure}
   \centerline{\includegraphics[width=.9\textwidth,clip=]{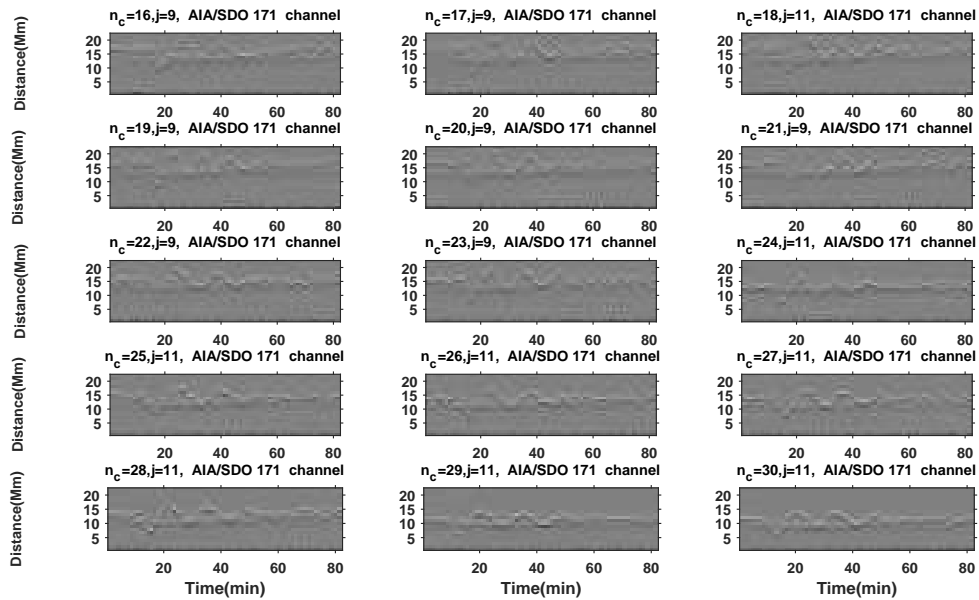}
              }\caption{  The same as in Figure 3 but for sectorial areas 16-30.}
           \vspace{+0.01\textwidth}
     \label{fig4}
   \end{figure}

\begin{figure}
   \centerline{\includegraphics[width=.9\textwidth,clip=]{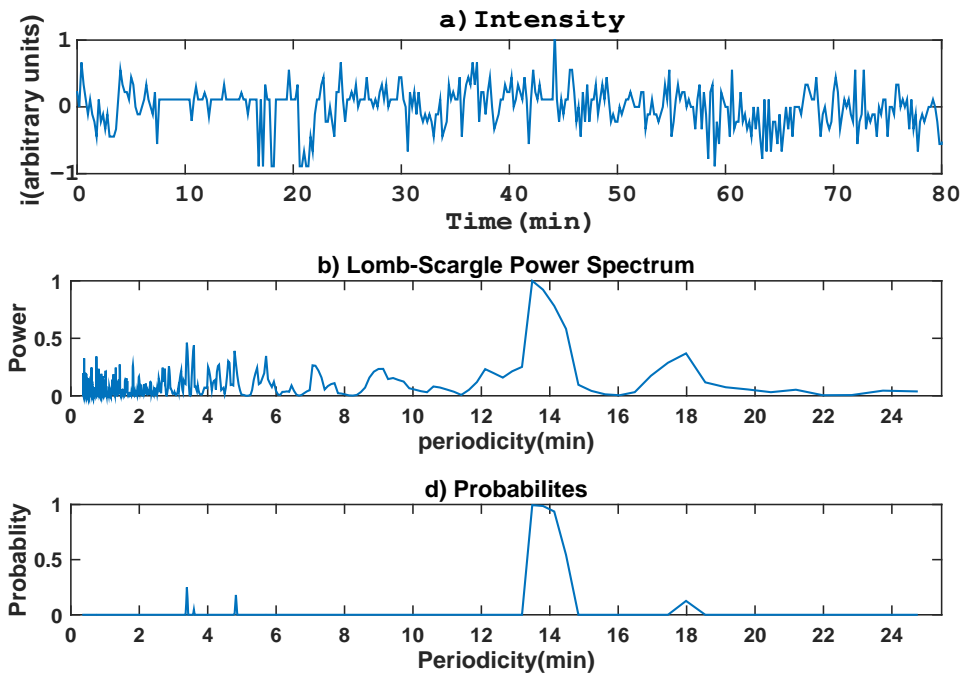}
              }\caption{a)  Temporal  variation of histogram-equalized and  background-subtracted intensity time series
              of macropixel (13,10)  are shown as function of time.
              b) Lomb-Scargle  power  spectral density of time series intensity  is shown as function of oscillation periods.
              c) The true-alarm probability of peaks in the Lomb-Scargle power spectra  are shown as function of oscillation periods.}
           \vspace{+0.01\textwidth}
     \label{fig5}
   \end{figure}

\begin{figure}
   \centerline{\includegraphics[width=.9\textwidth,clip=]{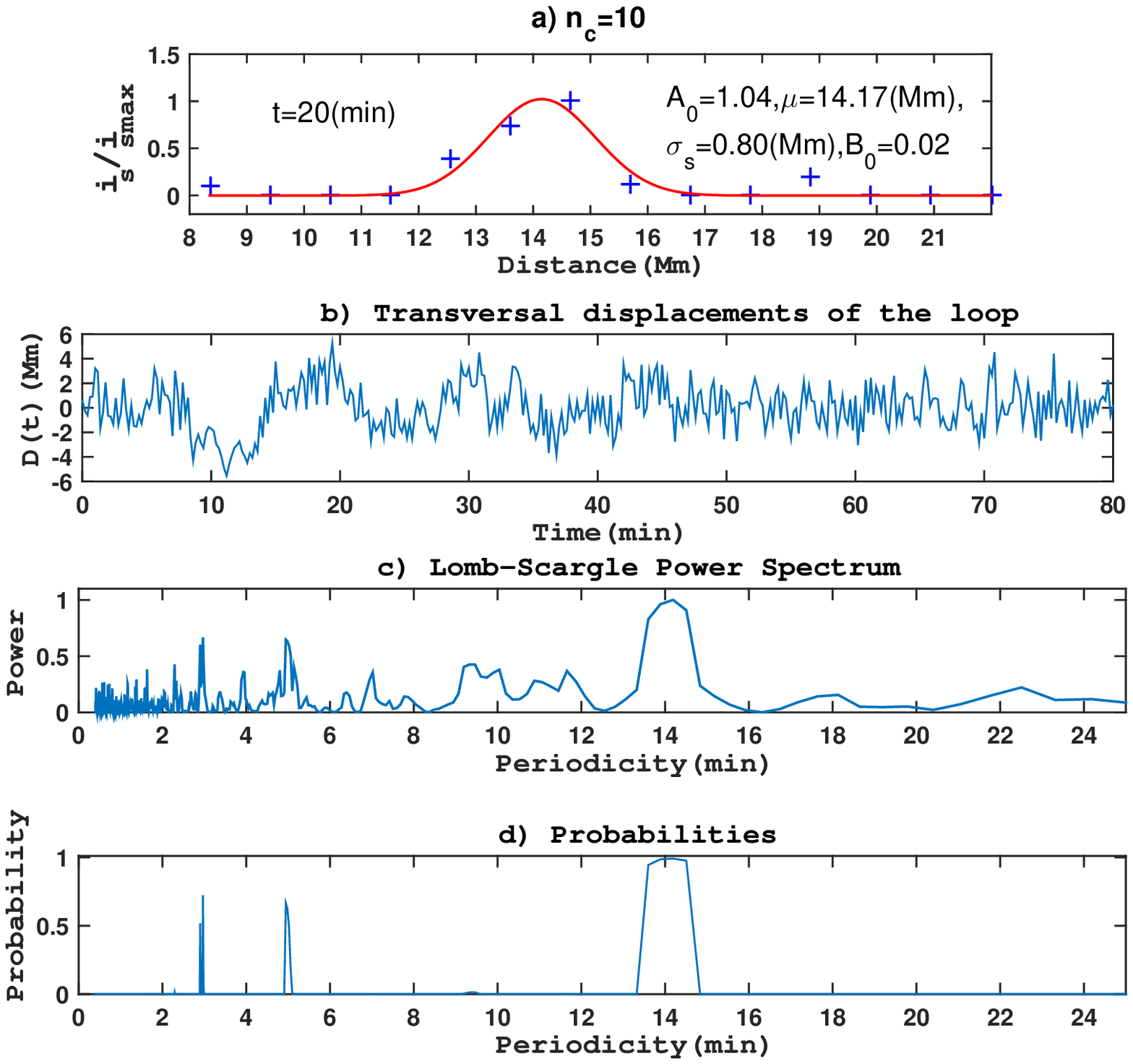}
              }\caption{a) Histogram-equalized,  background-subtracted  and  normalized intensity profile after  running-minimum subtraction ($ i_s(s,t)/i_{smax}(s,t))$ along the sectorial  10 at t=20 min are  plotted \textit{versus} the projected distance. This profile is fitted by an Gaussian function, $G(s,t) =A_0(t) \exp (-\frac{(s-\mu(t))^{2}}{2\sigma_s(t) ^{2}})+B_0(t)$ (\textit{red line}), and fit parameters are shown on the top part of the panel.
              b) Transversal displacements of the loop axis $ D(t)=(\mu(t)-<\mu(t)>$)  for sectorial 10  are shown as function of oscillation periods.
              c) Lomb-Scargle  power  spectral density of $D(t)$ is shown as function of oscillation periods.
              d) The true-alarm probability of peaks in the Lomb-Scargle power spectra  are shown as function of oscillation periods.}
           \vspace{+0.01\textwidth}
     \label{fig6}
   \end{figure}
 \section{Estimation of the physical parameters of  transversal coronal loop oscillation}
A coronal loop can be oscillated in various directions. A basic type of coronal loop oscillation is called transverse oscillation.
The transverse oscillation can be caused by sausage and kink type modes or it can be caused by the wake of a traveling disturbance  from nearby flares.
Here, the physical parameters of  transversal coronal loop oscillations  that  are powerful diagnostic tools for  coronal seismology are estimated.
\subsection{The measurement of  transversal coronal loop oscillation periods}
In this subsection, oscillation periods of transversal  motion of loop segments are estimated.
Transversal displacements of the loop have been interpreted in terms of the transverse
standing kink mode or the wake  of a traveling disturbance of nearby events,
since they trigger transversal displacements  in  the loop axis.
In the Figure \ref{fig5},  intensity time series, Lomb-Scargle  power  spectral and  probability  of   power  spectral of  macropixel (13,10) are displayed  in the panels a),  b) and c), respectively.
 Intensities  of macropixels   may be varied by the transversal and/or the longitudinal oscillations.
 So, in order to infer the  transversal coronal loop oscillations as function of time in each time$-$distance plots, a Gaussian function of the form
  \begin{eqnarray}\label{temp }
G(s,t) =A_0(t) \exp [-\frac{(s-\mu(t))^{2}}{2\sigma_s(t) ^{2}} ]+B_0(t),
\end{eqnarray}
is fitted to the histogram-equalized, background-subtracted  and normalized intensities along the paths by Matlab Statistics Toolbox Curve Fitting and Distribution Fitting.
This fitting yields the  peak of intensity profile $A_0(t)$, location of the loop axis $\mu(t)$,  standard deviation $\sigma_s(t)$,  and mean background intensity $ B_0(t) $  along a specific path in each time, respectively.
The oscillation periods of loop axis are calculated by Lomb-Scargle power spectral densities   of $ D(t)=(\mu(t)-<\mu(t)>$) time series.
For example, histogram-equalized,  background-subtracted  and  normalized intensity profile
 relative to the peak of intensity profile ($ i_s(s,t)/i_{smax}(s,t)$) for sectorial 10 at $t=20 $  minutes are  plotted versus the projected distance in the panel a) of Figure \ref{fig6}. The normalized intensity profiles (blue plus signs) is fitted by an Gaussian function (red line), and fit parameters are displayed in the top part of the panel.
Transversal displacements of the loop axis  ($D(t)$)  for sectorial 10   are shown as function of oscillation periods in the panel b) of Figure \ref{fig6}.
There are a number of techniques that allow to estimate the statistical significance of the
detected periodicity, for example the well-known   Lomb-Scargle method.
This  periodgram   algorithm gives  a better detection  accuracy and efficiency    for analysis of noisy time series  data.   Also, the probability of peaks  in the  power spectral density of noisy time series  data happening  by chance can be calculated. Furthermore,  this algorithm avoids  possible incorrect results that may occur from replacement of missing time  series data by interpolation methods.
For example, the  Lomb-Scargle  power spectral densities of  transversal displacements  for sectorial 10  are shown as function of oscillation periods in the panel c) of Figure \ref{fig6}.
Lomb-Scargle  power spectral densities  of transversal displacements show that there are indeed several statistically dominant peaks of power spectra in the range of P = $2-20$ minutes.
The true-alarm probability of significant peaks of power spectra  in the panel d)  show that only  $\rm{P}\approx 3, 5$ and 14 minutes are significant  oscillation period.
Comparisons of  probability of peaks before (panel c) of Figure \ref{fig5}) and after  running-minimum subtraction (panel d) of Figure \ref{fig6}) reveal that significant peaks are quite similar.
Furthermore, probability of significant peaks  in 30 different  sectorial areas
reveal  three or four significant oscillation  periods.
One dominant  peak  around $ \rm{P}= 14 $  minutes is present  in all Lomb-Scargle power spectral densities  of  sectorial areas. But, other peaks  are  completely  different in the power spectral densities.
\subsection{The measurement  of the damping of transversal loop oscillation}
Here, damping times of transversal loop oscillation are measured.
Also, the dependence of damping times on amplitude and also on its periodicity are investigated.
Lomb-Scargle  power spectral densities of   $D(t)$
show that spectral components of the transversal displacement  along the thirty different sectorial areas are
 significant at some specified oscillation periods in the range of
P = $2-20$ minutes. But, oscillation  periods around the $\rm{P}=14$ minutes
 are almost  dominant  in all power spectra.
Accordingly, the damping time of loop oscillation along the 30 different segments of coronal
loop are measured only for common dominant
oscillation periods.
So, in the Lomb-Scargle power spectral density  dominant components of loop axis displacement  at each sectorial areas are selected  by multiplying each of the  components  into the Gaussian filter of
\begin{eqnarray}\label{temp }
a^{'}_j =\sum_{i=1}^{N_f}a_i \exp [-\frac{(\nu_i-\nu_j)^{2}}{2\sigma_\nu ^{2}} ],
\end{eqnarray}
where the indices  $i$ and  $j$  are the rank of the harmonics,
$N_f$ is the  total number of the spectral components, $\sigma_\nu $  is standard deviation, $\nu_i$ and $\nu_j$ are  the frequency of the $i$th and $j$th harmonics,
$a_i $  and $a^{'}_j $  are  the  $i$th and $j$th  Fourier coefficient  before and after transformation, respectively.
 Two different kinds of theoretical damping mechanisms have been proposed for transversal coronal loop oscillation:
(1) The damping of  transversal coronal loop oscillations  may be produced by the dissipation mechanism.
Generally, resonant absorption is found to be the dominant damping mechanism compared to the other dissipation mechanism
(see, \textit{e.g.}  Aschwanden \textit{et al.}~\citeyear{Aschwanden2003}).
(2) The damping  may be produced by the wake of the traveling disturbance (see, \textit{e.g.} Rae and Roberts ~\citeyear{Rae1982}; Uralov ~\citeyear{Uralov2003}; Terradas, Oliver, and Ballester~\citeyear{Terradas2005}).
A localized disturbance from a solar flare  propagates toward the loop. As a consequence,  coronal loop axis  oscillates transversally at the external cutoff frequency, and it is damped  approximately with time as  $t^{-1/2}$  by the wake of the traveling disturbance.
To investigate the damping nature of  transversal coronal loop oscillation, the variation of  filtered displacement ($ D(t)$) with time are fitted by two different forms of damped sine function, $ \rm{f_1(t)= a_1 \exp(-(t-t_0)/\tau_d)\sin (2\pi (t-t_0)/P_1-\phi_{10})}$ and $\rm{ f_2(t)= a_2(t-t_0)^{-\alpha}\sin (2\pi (t-t_0)/P_2-\phi_{02})}$.
\begin{figure}
   \centerline{\includegraphics[width=.8\textwidth,clip=]{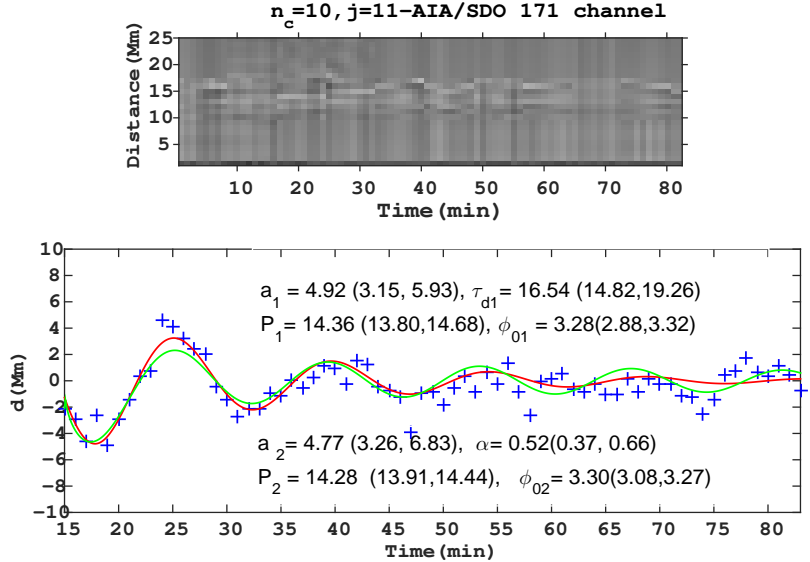}
              }\caption{Template time$-$distance plot for sectorial 10   is shown   in the  \textit{top panel}.
           The variation of  filtered transversal displacement  ($ D(t)$) with time for sectorial 10 at a  dominant oscillation period $\rm{P}=14$  min is shown as function of  time  in the bottom panel. Also,  amplitudes are fitted by two different forms of damped sine function, $\rm{ f_1(t)= a_1 \exp(-(t-t_0)/\tau_d)\sin (2\pi (t-t_0)/P_1-\phi_{01})}$ (red line), $\rm{ f_2(t)= a_2(t-t_0)^{-\alpha}\sin (2\pi (t-t_0)/P_2-\phi_{02})}$ (\textit{green line}), and fit parameters are shown on the top and bottom parts of panel, respectively.}
\vspace{+0.01\textwidth}
                \label{fig7}
\end{figure}
For example,  time$-$distance  plot  for sectorial 10  is  displayed  in the top panel of Figure \ref{fig7}. Also, the extracted amplitudes (blue plus signs) at a dominant oscillation period ($\rm{P}= 14$  min) are fitted by two different  forms of damped sine functions and fit parameters are written in the top and bottom part of  panel  of Figure \ref{fig7}, respectively.
\begin{table}
\small
\caption{Intensity profiles along 30 different sectorial areas  of the coronal loop for 400 images are fitted with two different  forms of damped sine function, $ \rm{f_1(t)= a_1 \exp(-(t-t_0)/\tau_d)\sin (2\pi (t-t_0)/P_1-\phi_{10})}$ and
$\rm{ f_2(t)= a_2(t-t_0)^{-\alpha}\sin (2\pi (t-t_0)/P_2-\phi_{02})}$. Magnitudes of fit  parameters, such as periodicity, damping time ($\rm{\tau_d}$),  damping quality ($\rm{\tau_d/P}$), magnitude values of $\alpha$  and phase angle ($\phi$)  of 30 sectorial areas that they are perpendicular to the visible loop strands observed in the  AIA/SDO 171 {\AA} passband.}
\begin{tabular}{|c|c|c|c|c|c||c|c|c|c|}
  \hline
  $\rm{n_c}$& $\rm{a_1}$  & $\rm{\tau_{d1}}$&   $\rm{P_1}$&    $\rm{\tau_{d1}/P_1}$&  $\rm{\phi_{01}}$& $\rm{a_2}$& $\rm{\alpha}$&   $\rm{P_2}$&   $\rm{\phi_{02}}$\\
   $ $& $\rm{(Mm)}$  & $\rm{(min)}$&   $\rm{(min)}$&  &  $\rm{(rad)}$& $\rm{(Mm)}$& $$&    $\rm{(min)}$&  $\rm{(rad)}$\\
  \hline
   1& 3.06 &  18.40&   13.86&  1.33 &  3.05&    3.23  &  0.55  &13.26   & 3.04\\
   2& 3.98 &  16.99&   14.59 &  1.17   & 3.19&    4.40&    0.69&  14.45&  3.14\\
   3& 3.86&  20.06&   13.02  &  1.54 & 3.04&    4.25&   0.59 &  13.31&    3.10\\
  4 & 3.42 &  21.46&   14.68&    1.460 & 3.32&    4.45&  0.67 & 14.05&   3.24\\
   5 & 3.35 &  18.58&   15.15 &  1.22 & 3.26&     4.40&   0.65&   15.13  &  3.17\\
  6 & 4.46&  16.67&   14.32& 1.16   & 3.22&    5.44&   0.75&  14.16 &   3.18\\
  7 & 4.64 &  19.65&   13.16  & 1.49& 3.07&     5.01&   0.62&   13.25 &   3.14\\
 8   & 5.10 &  18.09&   13.85 & 1.31& 3.09&      5.17 &   0.59&  14.27&   3.13\\
 9   & 5.11&   14.94&   14.43& 1.03& 3.21   & 5.14&   0.64&   15.03 &   3.14\\
 10  &  4.92&   16.54&   14.36  & 1.15& 3.29  &   4.77&    0.52 &  14.28  &  3.30\\
11  & 6.24&   13.52&   14.36  & 0.91& 3.13 &      7.58&   0.75  & 14.67&    3.14\\
12  & 6.32&  12.16&   14.14 & 0.86& 3.05&       7.44 &   0.72&   15.02&   3.12\\
13  &  6.76 &  15.58&   14.58  & 1.07& 3.09   &   6.70 &   0.60 &  14.27&    3.14\\
14  &  6.36 &  14.63&   15.20 &   0.96&  3.01&   6.57&   0.62 &  15.43 &   3.12\\
15  &  5.65 &  18.65&   15.73 &  1.18&  3.19 &    6.90 &   0.64&   15.75&    3.16\\
16  &  6.98&  15.70&   15.20&    1.03& 3.29&    6.45 &   0.63 &  14.49&    3.14\\
17  &  6.60 &  16.26&   14.33 &  1.13& 3.23 &   6.17 &   0.65&  12.99&    3.14\\
 18 &  -  &  -   &     -     &    - &    -& -  &   -&  -&    -\\
 19 &  -  &  -   &     -     &    - &    -& -  &   -&  -&    -\\
 20 &  5.64&   16.97&   12.29 &  1.38& 3.18&      6.18&    0.58 &  12.81&    3.15\\
 21 &  5.73 &  12.34&   12.38&  0.99& 3.12 &     5.32&   0.66 &  12.25&    3.14\\
22  &  5.73&   15.48&   13.97 &  1.11&  3.29 &     6.25 &  0.70 &  13.93 &   3.24\\
23  & 6.17 &   11.76&   12.25 &  0.87& 3.16&      6.24 &   0.67 &  12.12  &  3.14\\
24  &  6.80 &  15.00&   12.56 &  1.17&  3.22&   5.90 &   0.62 &  12.39 &   3.13\\
25  &  6.13&  17.63&   15.80 &   1.16&  3.33&    5.51 &   0.64&   15.13 &   3.14\\
26  &  6.16 &  16.30&   13.92&   1.17&  3.06&    5.74 &   0.65 &  14.99&    3.14\\
27  &  6.66 &  13.40  & 13.65&  0.98&  3.14 &      5.55&    0.61&   14.13 &   3.14\\
28  &  4.23&   16.39&   13.91&  1.18&  3.26&      3.20 &   0.51 &  13.16 &   3.14\\
29  &  4.44&  20.12&   14.51&   1.39& 3.23&      4.26  &  0.61&   14.20&    3.14\\
30  &  4.64&   20.29&   13.91&   1.46& 3.28 &     3.80&    0.63&   14.32&    3.14\\ \hline
\end{tabular}
\label{results2}
\end{table}
The extracted  physical quantities such as period of oscillation ($\rm{P}$), damping time ($\rm{\tau_d}$), damping quality ($\rm{\tau_d/P}$), phase angle ($\phi$) and magnitude values of $\alpha$, for filtered  background-subtracted intensity at a sequence of 400 images with a time cadence of 12 seconds  are listed in Table \ref{results2}  for 30 different segments  of the coronal loop.
The results of this analysis reveal that the magnitude  of damping time and damping quality  for dominant oscillation periods  of  segments  ($\rm{P}=12.25-15.80  $ ) along the coronal loop  are within the range of $ \rm{\tau_d}\simeq 11.76- 21.46~~\rm{min}$.
Oscillation of loop segments   attenuates with time roughly as $t^{-\alpha}$ that the magnitude  of $\alpha$
change from 0.55 to 0.75.
The damping quality of loop segments  are in the range of $ \rm{\tau_d/P}\simeq 0.86- 1.49$  which clearly indicates that  the damping  regime is strong for oscillations.
Also,  these results show  that  damping times  increase  with increasing  oscillation  period.
 Damping properties decrease  slowly with increasing amplitude, but they are not sensitive to the oscillation periods.
This range of oscillation periods and damping times of loop segments are in good agreement with previous findings (see, \textit{e.g.} Nakariakov \textit{et al.} ~\citeyear{Nakariakov1999}; Aschwanden \textit{et al.} ~\citeyear{Aschwanden2002};  Verwichte \textit{et al.}~\citeyear{Verwichte2004}; Wang and Solanki ~\citeyear{Wangs2004})
\subsection{The measurement of speed of  the flare-generated disturbance and the lower coronal eruption/ejection}
 Association of  transversal coronal loop oscillations with  a hypothetical driver of loop oscillations such as a flare, an LCE or a  CMEs  must be investigated  in order to identify a relationship between the excitation and damping of transversal coronal loop oscillations with these phenomena.
For this purpose, the possible solar flare, LCE, or CME associated with the
oscillating loop were checked  using Helioviewer, JHelioviewer, XRS/GOES,
LASCO/SoHO,  CME catalogue and  CACTus CME list.
By using these tools, a flare was found near the selected loop  at position $ \rm{(x_{flare},y_{flare})=(-960'',-192'')}.$
It   started around 2015 March 2, 09:09:00 UT and ended around 2015 March 2, 09:31:12 UT with
a peak flux of 1182.3 $\rm{erg/(cm^{2}s)}$ at 09:17:12 UT.
Also, an LCE  was detected  near the limb   at position $ \rm{(x_{LCE},y_{LCE})=(-1006.8'',-195.6'')}.$
This LCE  started around 2015 March 2, 08:21:07 UT and ended around 2015 March 2, 09:01:07 UT.
Moreover, in the vicinity of the loop, flare and  LCE, CME was not detected one hour before and after of these events start times.
Only one  CME  was found near the oscillating loop with LASCO/SoHO.
It  started around 2015 March 2, 12:09:05 UT  having
a radial velocity  269~$\rm{km~s^{-1}}$, an angular width 20 degree and a position angle 118 degree, respectively.
So, the speeds required for a hypothetical driver of transversal loop oscillations
to reach the oscillation loop segments  sites from the starting point  were estimated.
Following  Zimovets and  Nakariakov (\citeyear{Zimovets2015}), these  travel speeds  were found in the range of $\rm{v_{flare}=195-520\rm{km~s^{-1}}}$ and  $\rm{v_{LCE}=14-35\rm{km~s^{-1}}}$ by the ratio of the distances between the oscillating loop segments  and  sites of these events to the time differences  between the   starting times of LCE and flare with  the oscillation loop segments.
This result shows that oscillation  in a coronal loop  cannot be excited  by a  low-speed shock wave of flare,
 since the values of the Alfv\'{e}n speed  are   mostly more than $800~\rm{kms^{-1}}$
(see, \textit{e.g.} Nakariakov and Ofman~\citeyear{Nakariakov2001}; Stepanov \textit{et al.}~\citeyear{Stepanov2012};  De Moortel and Nakariakov~\citeyear{DeMoortel2012}; Zimovets and  Nakariakov \citeyear{Zimovets2015}).
\section{ Theoretical considerations }
Recently, space-based telescope observations  have shown that  transversal coronal loop oscillations are strongly damped, with damping times of only 1-2 oscillation periods.
Theoretically, two different  kinds of physical  mechanisms have been offered for damping of transversal coronal loop oscillations:  damping by  dissipation mechanism,  and  damping  by   the wake  of  a traveling  disturbance.
\subsection{ Dissipation mechanism }
Several dissipation  mechanisms  have been proposed for
 damping of transversal coronal loop oscillations.
Theoretical studies investigating the decay of transverse  oscillations of coronal loops
have concentrated on the effects of phase mixing, resonant absorption,
  lateral  and foot point wave leakage, gravitational
stratification and  magnetic field divergence.  Generally,
 resonant absorption is found
to be the  relevant dispassion mechanism for decay   of kink mode oscillations (see, \textit{e.g.} Aschwanden \textit{et al.} ~\citeyear{Aschwanden2003} and references therein).
The damping quality  of  a kink mode oscillating loop with a thin boundary layer due to  resonant absorption  is given by (see,\textit{e.g.} Goossens \textit{et al.}~ \citeyear{Goossens2002}; Ruderman and Roberts ~\citeyear{Ruderman2002};  Van Doorsselaere \textit{et al.}~\citeyear{VanDoorsselaere2004})
 \begin{eqnarray}
&&{ {\tau_d}/\rm{P}= C(\frac{r_{loop}}{l_{skin}})(\frac{1+\frac{\rho_e}{\rho_i}}{1-\frac{\rho_e}{\rho_i}})},
\label{fiting}
  \end{eqnarray}
where $\rm{C}$  is a  constant that depends on  the radial density profile,
 $\rm{r_{loop}}$ is the loop radius, $\rm{l_{skin}}$ is the skin depth of the coronal loop,
 $\rm{P}$ is the mode period ${2L/(2n-1)C_K}$, $n$ is harmonic number of the $n$th harmonic, $\rm{\rho_i}$ and $\rm{\rho_e}$ are the  internal and external plasma  density   of corona loop, respectively.
\subsection{ Wake of the traveling disturbance}
Uralov (\citeyear{Uralov2003}) and Terradas \textit{et al.}~(\citeyear{Terradas2005})  studied the effect of a localized disturbance from a solar flare on transverse motion of a loop.
In the zero-$\beta$ plasma limit, they found that dynamics of coronal loop and its environment is governed by the Klein-Gordon equation  with cutoff frequency $\omega_c=2\pi c_A /2L$
by assuming that  loop and its environment with  a uniform magnetic
field  is extended  between $z=\pm L/2$.
These authors  pointed out that transversal displacement of coronal loop  decays approximately
with time  as $t^{-1/2}$  by the wake of a traveling disturbance.
 The  wakefield model does not take into account the  field-aligned structuring of the corona, in other words, it does not have loops. But, the structuring could affect the wave behaviour quite significantly, as it was shown by Yuan \textit{et al.}~(\citeyear{Yuan2015}).
Decay  of transversal loop oscillations due to the wake of a  disturbance
is not related to any dissipation mechanism such as  resonant absorption, phase
mixing,  \textit{etc}. Also, these transversal displacement of coronal loops
are not necessarily related to  the kink mode oscillations, although standing kink oscillations   can occur
after the decay process  by means of  the wave packet energy.
\section{ Discussion and Conclusions}
In this paper, it is attempted to provide more insight into
the excitation, damping and  temporal evolution of  transversal displacements of the loop
axis  near flares, and  also an attempt is made to deduce a quantitative relation between oscillation period, damping time,  damping quality, oscillation amplitude,  dissipation mechanism, lower coronal eruption of some unstable coronal magnetic field configuration and the wake phenomena that is produced by  a traveling disturbance in solar coronal loops.
For this purpose, the nature of transversal oscillation of a coronal loop  ( as observed  with AIA/SDO) is analyzed  by the histogram-equalization  method and Lomb-Scargle periodgram algorithm method  that  reveal a clearly better detection accuracy and efficiency  for analysis of  noisy time series data.
The  values  of physical parameters of transversal corona loop segments are  extracted
in the  171 $\rm {\AA}$ passband.
Observational results obtained can be summarized in the following main points:
\begin{description}
\item[$\bullet$]
The observational results of this study show  that histogram equalization technique  has  enhanced
the contrast of  transversal coronal loop oscillations in the time$-$distance plots (s ee Figures \ref{fig3} and \ref{fig4}) compared
to the previous studies (see, \textit{e.g.} Aschwanden and Schrijver~\citeyear{Aschwanden2011}; Goddard \textit{et al.}~\citeyear{Goddard2016585}).
 \item[$\bullet$] The values of the physical parameters  of transversal displacements of the loop segments  that have been estimated in this  analysis
            are in good agreement with previous findings (see, \textit{e.g.}  Nakariakov \textit{et al.} ~\citeyear{Nakariakov1999};
      Aschwanden and  Schrijver ~\citeyear{Aschwanden2002}; Aschwanden \textit{et al.}~\citeyear{Aschwanden2002}; Verwichte \textit{et al.}~\citeyear{Verwichte2004}; Wang and Solanki ~\citeyear{Wangs2004};
     Goddard and Nakariakov \citeyear{Goddard2016590}).
\item[$\bullet$] The Lomb-Scargle algorithm shows that spectral components of the transversal displacement  along the 30 different sectorial areas measured are dominant  at some specified oscillation period in the range of  P =$ 2-20$ minutes. But, probability of  the observed  peaks in the power spectral densities reveal that  oscillation  periods around the $\rm{P}=14$ minutes  are the most  dominant component in all power spectral densities of segments.
\item[$\bullet$] The magnitude  of  damping times and damping qualities of  the dominant oscillation period of loop segments ($\rm{ P=12.25-15.80}$ minutes) that  were determined for the exponential damping
      are measured in the range of $\rm{ \tau_d=11.76-21.46}$  minutes and $\rm{\tau_d/P=0.86-1.49},$ respectively.
 \item[$\bullet$]
The magnitude of  damping qualities clearly indicate that damping regime is strong for oscillations.
Also, the observational results of this analysis reveal that damping qualities decrease  slowly with increasing  amplitude of the oscillation, but they are not sensitive  to the  oscillation periods.
\item[$\bullet$]
Oscillation of loop segments attenuate with time roughly as that $t^{-\alpha}$  values of $\alpha$ for 30 different segments change  from 0.51 to 0.76. These values are  in good agreement  with theoretical predictions. But, the estimated  probability of peaks in Lomb-Scargle power spectral densities of  oscillation segments reveal  that there is indeed only one dominant peak in each power spectral density.
So, considering that the damping of  transversal coronal loop oscillation  is induced  by the wake of a traveling disturbance is not physically compatible with the dissipation mechanism.
\item[$\bullet$]
Finally, considering that decaying kink oscillations are excited  by a
lower coronal eruption of some unstable configuration of coronal magnetised plasma is acceptable as it is compatible with a blast shock wave due to a nearby flare.
 \end{description}
  \begin{acks}
The author thanks the anonymous referee for the very helpful comments and suggestions.
This work was carried out with the support of the University of Qom.\\
 \end{acks}
 \textbf{Disclosure of Potential Conflicts of Interest} The authors declare that they have no conflicts of interest.

\end{article}
\end{document}